\documentclass[twocolumn,aps,prl,floatfix,showpacs]{revtex4}
\usepackage[latin1]{inputenc}
\usepackage{graphicx}
\usepackage{amssymb}

\makeatletter

\begin{document}

\title{Decoherence by Correlated Noise and Quantum Error Correction}

\author{E.~Novais}

\affiliation{Department of Physics, Duke University, Durham NC 27708-0305}

\author{Harold~U.~Baranger}

\affiliation{Department of Physics, Duke University, Durham NC 27708-0305}

\date{July 19, 2006, in press Phys. Rev. Lett. {\bf 97}, 040501 (2006)}

\begin{abstract}
We study the decoherence of a quantum computer in an environment which is inherently correlated in time and space. We first derive the nonunitary time evolution of the computer and environment in the presence of a stabilizer error correction code, providing a general way to quantify decoherence for a quantum computer. The general theory is then applied to the spin-boson model. Our results demonstrate that effects of long-range correlations can be systematically reduced by small changes in the error correction codes.
\end{abstract}


\maketitle

Quantum computers bear the promise to solve certain problems exponentially faster than their classical counterparts \cite{NC00}. 
Although small computers have been successfully tested \cite{CVZ+98}, the development of large computers has been hindered by decoherence. The most promising method to tame decoherence is quantum error correction~(QEC) \cite{NC00,CVZ+98,Ste96,KLV00}. In QEC, it is usually assumed that correlations in the environment either are non-existent or decay exponentially in time and space. In contrast, recent work argues that correlated environments can lead to quadratically worse error-levels \cite{AHH+02,K04,Ter,AGP}. Though the assumption of uncorrelated noise is often reasonable, it is not fulfilled in several physical systems proposed for realizing quantum computers, notably solid state systems using superconductors \cite{SMS02} or quantum dots \cite{ALS02}. Hence, it is far from clear how much protection from decoherence QEC gives in these important cases \cite{AHH+02,Ter}.

In this paper, we consider the long time dynamics of a quantum computer immersed in a correlated quantum environment and protected by QEC.  
First, we describe the parameters which quantify the level of protection from QEC
and give explicit expressions for them.  Second, we calculate these quantities in a concrete example: the spin-boson model \cite{LCD+87}.  This model is directly applicable to solid state quantum computers \cite{SMS02,ALS02} but formally outside the scope of QEC \cite{Ter}.

Our work shows that some protection against long-range correlations can be built into QEC codes.  The new element here is that the periodic measurements in the QEC method separate the environmental modes into high and low frequencies.  This natural ``new'' scale can then be used to engineer quantum codes to better cope with the long-range correlations.

To follow the long time behavior of the computer, we remove non-essential elements and assume: (1) Quantum gates are perfect and operate much more quickly than the characteristic response of the environment. (2) States of the computer can be prepared with no errors. (3) Thermal fluctuations 
are
suppressed. Finally, for clarity in the spin-boson example, we consider ohmic coupling between the environment and the qubits. Extensions to sub-ohmic and super-ohmic coupling are straightforward.

\emph{Decoherence, QEC, and correlations---}The wave functions of the computer and the environment are unavoidably entangled during their time evolution. When a measurement is performed, this entanglement is translated into the probability of an error with respect to the ideal state of the computer. This general feature of open quantum systems is known as decoherence. In order to reduce its effects, QEC encodes the computer state in a subspace of a larger Hilbert space, $\mathcal{H}$, such that by measuring observables in $\mathcal{H}$ the wave functions of the computer and the environment are disentangled to some order in $\epsilon$, the probability of a single error in a QEC cycle.  The outcome of the measurement, called the syndrome, discriminates among the possible errors in the computer, allowing a correction to be  made.
Note that by performing a measurement on the computer, the environment's dynamics is restricted; thus, it is crucial to follow the non-unitary evolution of the entire system.

The difficulty with correlated environments is that $\epsilon$ is no longer time independent \cite{AHH+02,K04,Ter,AGP}: past errors or gates can substantially change $\epsilon (t)$. Consider, for instance, that at time $t_{1}$ an error occurs in the computer. Later, at time $t_{2}$, this is realized and corrected.  Although the error was detected, its precise time is unknown; therefore, when calculating the probability of a future error, all possible events at $t<t_{2}$ must be taken into account.  If the environmental correlations decay algebraically, there is no characteristic scale at which to truncate this sum.
Hence, it may be hard to estimate $\epsilon (t)$.
Sometimes it is sufficient to evaluate upper bounds on the error \cite{KLV00}; however, these estimates may be poor for certain models.
For instance, 
correlated
noise was studied in \cite{Ter}, which derived an upper bound on the error strength in the Ohmic spin-boson
model.
This bound is linear in the bosonic ultraviolet cut-off, $\Lambda$, which is the only scale in the model.
One of our main points is that the dynamics imposed by the QEC code provides another scale, which can then be used to generate better codes and/or bounds.
We specifically discuss the spin-boson case, but our results can be adapted to other situations.

\emph{Time evolution of encoded qubits---}The environment usually can be described by a Hamiltonian, $H_{0}$. Although other couplings between the computer and the environment are possible, we focus on the case of local vectorial coupling: $V=\frac{\lambda}{2}\sum_{{\mathbf{x}}}\vec{f}\left({\mathbf{x}}\right)\cdot\vec{\sigma}\left({\mathbf{x}}\right)$, where $\vec{\sigma}\left({\mathbf{x}}\right)$ are Pauli matrices for the qubits, $\lambda$ is the coupling strength, and $\vec{f}\left({\mathbf{x}}\right)$ is a function of environment operators.

QEC is essentially perturbative in $V$. It is therefore natural to define the interaction representation: $\hat{V}\left(t\right)=e^{\frac{i}{\hbar}H_{0}t}Ve^{-\frac{i}{\hbar}H_{0}t}$.  In this representation, the time evolution of the system during a QEC cycle that starts at time $t=0$ and ends at time $t=\Delta$ is
\begin{equation}
\hat{U}\left(\Delta,0\right)  =  T_{t}
e^{-i\frac{\lambda}{2}\sum_{{\mathbf{x}}}\int_{0}^{\Delta}dt\vec{f}\left({\mathbf{x}},t\right)\vec{\sigma}\left({\mathbf{x}}\right)}
\label{eq:unitaryevolution}
\end{equation}
where $T_{t}$ is the time ordering operation. It is also simple to describe gates that are done faster than the environmental response. If a gate $R_{\tau}$ is executed at time $\tau<\Delta$, at the end of the QEC cycle the time evolution is given by $\hat{U}\left(\Delta,\tau\right)R_{\tau}\hat{U}\left(\tau,0\right)$.

At time $t=\Delta$, the measurement selects a set of terms from the r.h.s. of Eq.~(\ref{eq:unitaryevolution}). Only this set is carried to the next QEC cycle. For stabilizer codes \cite{NC00}, it is straightforward to identify which operators must be kept. First, because the measurements are performed on individual logical qubits, it is sufficient to analyze the $n$ physical qubits that define a logical one. All unitary operations in the Hilbert space of these $n$ qubits can be written using the Pauli group $G_{n}$.  The subgroup $E$ of $G_{n}$ of all possible errors in the computer is called the error set. Intuitively, $E$ is given by the Pauli matrices appearing in each term of an expansion of the the r.h.s. of Eq.~(\ref{eq:unitaryevolution}). For instance, the element $g=\sigma^{x}\left(\mathbf{x}_{1}\right)\sigma^{z}\left(\mathbf{x}_{2}\right)\in E$ is given by the time-ordered term $\lambda^{2}\int_{0}^{\Delta}dt_{1}\int_{0}^{t_{1}}dt_{2}f^{x} \left({\mathbf{x}_{1}},t_{1}\right)f^{z}\left({\mathbf{x}_{2}},t_{2}\right) \sigma^{x}\left({\mathbf{x}_{1}}\right)\sigma^{z}\left({\mathbf{x}_{2}}\right) $.

The next step is to decompose $E$ according to the possible values of the syndrome.  This is done by noticing that the measurement will not distinguish elements inside the logical Hilbert space. Thus, the logical Pauli Group $\bar{G}_{1}$, generated by the identity and the logical Pauli matrices, $\left\{ I,\bar{X},\bar{Y},\bar{Z}\right\} $, defines the subgroup $E_{0}=\bar{G_{1}}\cap E$. The partition $\texttt{P}$ of $E$ given by all left cosets of $E_{0}$ in $E$ sorts the errors by their syndromes. Using this fact, the r.h.s. terms of Eq.~(\ref{eq:unitaryevolution}) can be re-ordered as $U\left(\Delta,0\right)=\sum_{m}u_{m}\left(\Delta,0\right)$ with respect to $\texttt{P}$. Finally, a QEC code defines an appropriate recovery operation for each syndrome, $\mathtt{R}=\left\{ r_{m}\right\} $.  At the time of the measurement, $t=\Delta$, only one $\left\{ u_{m}\right\} $ is selected and the corresponding operation, $r_{m}\in\mathtt{R}$, is performed. The overall time evolution for a QEC cycle is $\upsilon_{m}\left(\Delta,0\right)=r_{m}\left(\Delta\right)u_{m}\left(\Delta,0\right)$; thus, interference between terms with different syndromes is eliminated. 

The analogous result for many logical qubits and a sequence of $N$ QEC cycles follows directly from the above:
\begin{equation}
\Upsilon_{{\mathbf{w}}}  =  \upsilon_{w_{N}}\big(N\Delta,\left(N-1\right)\Delta\big)...\upsilon_{w_{1}}\big(\Delta,0\big)
\label{eq:defUpsilon}
\end{equation}
where ${\mathbf{w}}$ is the particular history of syndromes for all the qubits and $\upsilon_{w_i}$ is simply the product of evolutions for the individual qubits. Each history comes with the associated probability
\begin{equation}
\mathcal{P}\left(\Upsilon_{{\mathbf{w}}}\right)  =  \left\langle \varphi_{0}\right|\left\langle \psi_{0}\right|\Upsilon_{{\mathbf{w}}}^{\dagger}\Upsilon_{{\mathbf{w}}}\left|\psi_{0}\right\rangle \left|\varphi_{0}\right\rangle \label{eq:hisprob}
\end{equation}
where $\varphi_{0}$ and $\psi_{0}$ are the initial states of, respectively, the environment and the encoded qubits. Finally, QEC only partially disentangles the environment and the qubits. There is always some residual decoherence, which can be found from the reduced density matrix
\begin{equation}
\rho_{\vec{r},\vec{s}}\left(\Upsilon_{{\mathbf{w}}}\right)  =  \frac{\left\langle \varphi_{0}\right|\left[\left\langle \psi_{0}\right|\Upsilon_{{\mathbf{w}}}^{\dagger}\left|\vec{s}\right\rangle \left\langle \vec{r}\right|\Upsilon_{{\mathbf{w}}}\left|\psi_{0}\right\rangle \right]\left|\varphi_{0}\right\rangle }{\left\langle \varphi_{0}\right|\left\langle \psi_{0}\right|\Upsilon_{{\mathbf{w}}}^{\dagger}\Upsilon_{{\mathbf{w}}}\left|\psi_{0}\right\rangle \left|\varphi_{0}\right\rangle }\label{eq:rdmlqubit}
\end{equation}
with $\vec{r}$,$\vec{s}$ elements of the logical subspace.

Eqs.~(\ref{eq:defUpsilon})-(\ref{eq:rdmlqubit}), our central formal result, can be used to assess the protection offered by a QEC code. They show that $\Delta$ is a natural scale in the field theory that describes the system's evolution. Although the consequences of this imposed scale are model dependent, its existence can be used to construct better QEC codes; to show this explicitly, we now turn to an example.

\emph{Decoherence in the spin-boson model---}The spin-boson model deals with generic two state systems coupled linearly with an infinite set of harmonic oscillators. In its general form, the model includes a finite tunneling amplitude and a bias between the two states. For qubits, these two features model ``imperfections'' and so are not considered here; they are not fundamental to the understanding of decoherence in a correlated environment \cite{Unr95}. As we consider only ``perfect'' qubits, the computer experiences errors only due to dephasing. Finally, we choose to consider the case of linear coupling to an ohmic bath.  We stress that these two choices do not restrict our results, but allow for a convenient notation.

The Hamiltonian of the model can be written as
\begin{equation}
H  \hspace{-0.04in}=  \hspace{-0.05in}\frac{v_{\rm b}}{2}\hspace{-0.03in}\int_{-\infty}^{\infty}\hspace{-0.15in}dx\left[\partial_{x}\phi\left(x\right)\right]^{2}\hspace{-0.06in}+\left[\Pi\left(x\right)\right]^{2}\hspace{-0.0in}+\sqrt{\frac{\pi}{2}}\lambda\sum_{n}\partial_{x}\phi\left(n\right)\sigma_{n}^{z},
\label{eq:H0+V}
\end{equation}
where $\phi$ and $\Pi=\partial_{x}\theta$ are canonical conjugate variables, $\sigma_{n}^{z}$ act in the Hilbert space of the qubits, $v_{\rm b}$ is the velocity of the bosonic excitations, and $\hbar=k_{B}=1$. The bosonic modes have an ultraviolet cut-off, $\Lambda$, that defines the short-time scale of the field theory, $t_{\rm uv}=\left(\Lambda v_{\rm b}\right)^{-1}$.  Following our general assumption, we regard gates as perfect and with operation time $t_{g}\ll t_{\rm uv}$.

Between gates, the exact time evolution of a qubit in the interaction picture can be expressed as the product of two vertex operators of the free bosonic theory, $U_{n}\left(t,0\right)  =  e^{i\sqrt{\frac{\pi}{2}}\lambda \left[\theta\left(n,t\right)-\theta\left(n,0\right)\right] \sigma_{n}^{z}}$, and a coherent evolution that is irrelevant to our discussion \cite{Unr95}.  Hence, it is straightforward to express the reduced density matrix as a bosonic correlation function. In particular, for a single qubit in the initial state $\left|\psi_{0}\right\rangle \!= \! \alpha\left|\uparrow\right\rangle +\beta\left|\downarrow\right\rangle$, the off-diagonal element is
\begin{equation}
\rho_{\uparrow\downarrow}^{\left(n\right)}\left(\Delta\right)  =  \alpha\beta^{*}\left\langle 0\right|e^{\sqrt{2\pi} i\lambda\left[\theta\left(n,\Delta\right)-\theta\left(n,0\right)\right]}\left|0\right\rangle =\alpha\beta^{*}e^{-\epsilon}
,\label{eq:red1}
\end{equation}
where $\left|0\right\rangle $ is the bosonic vacuum, $\epsilon \!=\! \lambda^{2}\ln\big[1+\left(\Lambda v_{b}\Delta\right)^{2}\big]/2$ is the probability of an error, and $\Lambda v_{b}\Delta\gg1$.  

\begin{figure}[tb]
\includegraphics[%
  width=0.80\columnwidth]{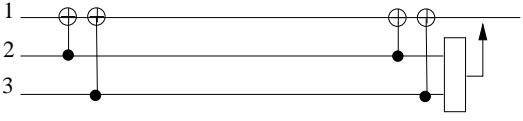}
\caption{\label{cap:QEC}Steane's 3 qubit quantum error correction (QEC) code \cite{NC00}. The initial wave function, 
$\left|\psi_{0}\right\rangle\otimes 
( \left|\uparrow\right\rangle \!+\! \left|\downarrow\right\rangle )/2
\otimes
( \left|\uparrow\right\rangle \!+\! \left|\downarrow\right\rangle )/2$,
is encoded by two controlled-NOT (CNOT) gates,
$R_{\rm CNOT} = \sigma_{i}^{-}\sigma_{i}^{+}\sigma_{j}^{x} \!+ \sigma_{i}^{+}\sigma_{i}^{-}$,
into an entangled state $\left|\psi_{\rm encode}\right\rangle =\alpha\left|\bar{\uparrow}\right\rangle + \beta\left|\bar{\downarrow}\right\rangle $
with $\left|\bar{\uparrow}\right\rangle =\left(\left|\uparrow\uparrow\uparrow\right\rangle +\left|\uparrow\downarrow\downarrow\right\rangle +\left|\downarrow\uparrow\downarrow\right\rangle +\left|\downarrow\downarrow\uparrow\right\rangle \right)/2$
and $\left|\bar{\downarrow}\right\rangle =\left(\left|\downarrow\downarrow\downarrow\right\rangle +\left|\downarrow\uparrow\uparrow\right\rangle +\left|\uparrow\downarrow\uparrow\right\rangle +\left|\uparrow\uparrow\downarrow\right\rangle \right)/2$.
After some time, the information is decoded by a second pair of CNOT gates.  An error
in $\left|\psi\right\rangle $ is identified by measuring the the value of
$\sigma_{2}^{x}$ and $\sigma_{3}^{x}$ (rectangle). The cycle of QEC ends with the
correction of a possible phase-flip (arrow). } 
\end{figure}

\emph{QEC and the spin-boson model---}The simplest QEC protocol is Steane's three qubit code (see Fig.~\ref{cap:QEC}). It is designed to protect a logical qubit from a single phase-flip, that is from dephasing in lowest order in the coupling to the environment.  Thus, we illustrate our discussion of QEC in correlated environments by applying this code to three qubits that dephase according to Eq.~(\ref{eq:H0+V}).

The error set of a single logical qubit is $E=\{  I,\,\sigma_{j}^{z},\,\sigma_{j}^{z}\sigma_{k\neq j}^{z},\,\sigma_{1}^{z}\sigma_{2}^{z}\sigma_{3}^{z}\} $, where $j,k=\left\{ 1,2,3\right\} $.  The logical Pauli matrix $\bar{Z}=\sigma_{1}^{z}\sigma_{2}^{z}\sigma_{3}^{z}$ is contained in $E$. Thus, the subgroup $E_{0}=\left\{  I,\,\bar{Z}\right\} $ can be used to partition the error set into four equivalence classes, each with its respective recovery operation, $\texttt{P}\leftrightarrow\texttt{R}$:
\begin{eqnarray}
\left\{  I,\,\bar{Z}\right\} \leftrightarrow I\,\,\,, &  & \left\{ \sigma_{1}^{z},\,\sigma_{2}^{z}\sigma_{3}^{z}\right\} \leftrightarrow\sigma_{1}^{z}, \nonumber\\
\left\{ \sigma_{2}^{z},\,\sigma_{1}^{z}\sigma_{3}^{z}\right\} \leftrightarrow\sigma_{2}^{z}, &  & \left\{ \sigma_{3}^{z},\,\sigma_{1}^{z}\sigma_{2}^{z}\right\} \leftrightarrow\sigma_{3}^{z}.
\end{eqnarray}

The first class, $\{  I,\,\bar{Z}\} $, corresponds to a superposition of the state where no error occurred, $\left|\psi_{\rm encode}\right\rangle $, and that with a logical phase flip, $\bar{Z}\left|\psi_{\rm encode}\right\rangle $. Similarly, the other three classes are superpositions of states with one and two phase-flips on physical qubits. When the recovery operations are done, the logical qubit is once again a superposition of $\left|\psi_{\rm encode}\right\rangle $ and $\bar{Z}\left|\psi_{\rm encode}\right\rangle $. Although the final state is seemingly the same, each possible evolution has a different bosonic content.

The bosonic parts are written compactly by separating the time evolution operator into ``even'' and ``odd'' orders: $U_{j}\left(\Delta,0\right)=\eta_{j} \!+\! i\nu_{j}\sigma_{j}^{z}$, where $\eta_{j} = \cos\left[\sqrt{\frac{\pi}{2}}\lambda\left[\theta\left(j,t\right)\!-\!\theta\left(j,0\right)\right]\right]$, $\nu_{j}=\sin\left[\sqrt{\frac{\pi}{2}}\lambda\left[\theta\left(j,t\right)\!-\!\theta\left(j,0\right)\right]\right]$.  Thus, the time evolution by the end of a QEC cycle is
\begin{eqnarray}
\upsilon_{0}\left(\Delta,0\right) & = & \eta_{1}\eta_{2}\eta_{3}I-i\nu_{1}\nu_{2}\nu_{3}\bar{Z}, \nonumber\\
\upsilon_{1}\left(\Delta,0\right) & = & i\nu_{1}\eta_{2}\eta_{3}I-\eta_{1}\nu_{2}\nu_{3}\bar{Z},\\
\upsilon_{2}\left(\Delta,0\right) & = & i\eta_{1}\nu_{2}\eta_{3}I-\nu_{1}\eta_{2}\nu_{3}\bar{Z},\quad\textrm{or} \nonumber\\
\upsilon_{3}\left(\Delta,0\right) & = & i\eta_{1}\eta_{2}\nu_{3}I-\nu_{1}\nu_{2}\eta_{3}\bar{Z}. \nonumber
\end{eqnarray}

Given a particular history of syndromes, ${\mathbf{w}}$, it is straightforward to find both its likelihood, $\mathcal{P}\left(\Upsilon_{{\mathbf{w}}}\right)$, and the corresponding contribution to the off-diagonal element of the reduced density matrix, $\rho_{\bar{\uparrow}\bar{\downarrow}}\left(\Upsilon_{{\mathbf{w}}}\right)$. For the moment, we consider the case of qubits separated by a distance much larger than $v_{\rm b}N\Delta$, so that spatial correlations can be disregarded. Although a very particular example, this is illustrative of the general discussion.

By the end of the first QEC cycle, there are two different cases to consider:
$\Upsilon_{{\mathbf{w}}}=\upsilon_{0}$ with probability
$\mathcal{P}\left(\upsilon_{0}\right)\simeq1-3\epsilon/2$ and
$\Upsilon_{{\mathbf{w}}}=\upsilon_{\left(1,2,3\right)}$ with
$\mathcal{P}\left(\upsilon_{\left(1,2,3,\right)}\right)\simeq\epsilon/2$.  Each one causes an intrinsic dephasing for the logical qubit
\begin{eqnarray}
\rho_{\bar{\uparrow}\bar{\downarrow}}\left(\upsilon_{0}\right) = \alpha\beta^{*}\frac{3e^{-\epsilon}+e^{-3\epsilon}}{1+3e^{-2\epsilon}}\simeq\alpha\beta^{*}\left(1-\frac{\epsilon^{3}}{4}+...\right),\nonumber \\
\rho_{\bar{\uparrow}\bar{\downarrow}}\left(\upsilon_{\left(1,2,3\right)}\right) = \alpha\beta^{*}e^{-\epsilon}\simeq\alpha\beta^{*}\left(1-\epsilon+...\right)
\;.\quad \label{eq:cyclewitherr}
\end{eqnarray}
Note that by the end of a cycle with an ``error'', the logical qubit has suffered precisely the same decoherence as for an unprotected physical qubit, Eq.~(\ref{eq:red1}). A cycle with such decoherence has, however, become a ``rare'' event; this exemplifies the benefit provided by QEC.

If there were no correlations (or in this case memory) between QEC cycles, then the likelihood of an error would be history independent and given by (\ref{eq:cyclewitherr}). In this uncorrelated limit, fault-tolerance can then be proved \cite{NC00,Ste96}.

\emph{Correlation between QEC cycles---}We now address to what extent correlation between cycles changes the likelihood of errors.  In the current example, the contribution to $\mathcal{P}\left(\Upsilon_{{\mathbf{w}}}\right)$ due to long-range correlation is elegantly evaluated by the Operator Product Expansion (OPE) \cite{DFM97}. The idea is to express the evolution of the environment during each cycle as a single operator that will capture the most relevant contributions to the correlation functions that need to be evaluated.

In order to use the OPE, $\mathcal{P}\left(\Upsilon_{{\mathbf{w}}}\right)$ must be re-ordered such that operators with arguments closer in space and time are put together.  Hence, in the spin boson case, all local operators are products of $\eta_{j}^{2}$ and $\nu_{j}^{2}$. The OPE decomposes $\eta_{j}^{2}$ and $\nu_{j}^{2}$ into high frequency, $\Delta^{-1}<\omega<v_{\rm b}\Lambda$, and low frequency, $\omega<\Delta^{-1}$, parts. The high frequencies are integrated out and the most relevant components of the low frequencies are kept:
\begin{eqnarray}
A_{\pm} & \sim & \left\{1\pm
e^{-\epsilon} :\!\cos\left[\sqrt{2 \pi} \lambda\partial_{t}
\theta\left(j,0\right)\Delta
\right]
\!:\right\}/2\;,\label{Apm}
\end{eqnarray}
where $:\,:$ represents normal ordering, $A_{+} \!= \eta_{j}^{2}$ and $A_{-}=\nu_{j}^{2}$.  In essence, this procedure captures the long range effects on the environment of the dynamics imposed by the QEC code by using operators which have the ``new'' ultraviolet cut-off $\Delta^{-1}$.  The leading order effects of correlations in $\mathcal{P}\left(\Upsilon_{{\mathbf{w}}}\right)$ are given by the effective operators 
\begin{eqnarray}
\upsilon_{0}^{2} \sim  1 - 3 \epsilon/2 - \sum_{j=1}^{3}\frac{\pi\lambda^{2}\Delta^{2}}{2}
:\!\left[\partial_{t}\theta\left(j,0\right)\right]^{2}\!: \;, 
\label{OPE1} \\ 
\upsilon_{j=\left\{ 1,2,3\right\} }^{2}  \sim  \epsilon/2 + \frac{\pi \lambda^{2}\Delta^{2}}{2}:\!\left[\partial_{t}\theta\left(j,0\right)
\right]^{2}\!: \;.
\label{OPE2}  
\end{eqnarray}
Hence, the probability, $\mathcal{P}\left(\Upsilon_{{\mathbf{w}}}\right)$, of a history ${\mathbf{w}}$ is easily evaluated using Wick's theorem and the fact that $\langle:\!\left[\partial_{t}\theta\left(j,t\right)\right]^{2}\!:\;
:\!\left[\partial_{t}\theta\left(j,0\right)\right]^{2}\!:\rangle \simeq 1/\left(2 \pi^{2} t^{4}\right)$.

The simplest case is to calculate the probability of having errors in the QEC cycles starting at times $t_{1}$ and $t_{2}$. From Eqs.~(\ref{OPE1}) and (\ref{OPE2}), it is straightforward that, in leading order in $\lambda$, $\mathcal{P}\left(...\upsilon_{j}....\upsilon_{j}...\right) \approx\left(\epsilon/2\right)^{2}+\lambda^{4}\Delta^{4} /[8\left(t_{1}-t_{2}\right)^{4}]$, where the first term is the uncorrelated probability and the second is due to correlations between errors in different cycles.  Therefore, for $N \ll 1/ \lambda^{2}$ QEC cycles, the probability of having two errors of any kind is $\mathcal{P}_{2}\approx[\left(\epsilon/2 \right)^{2}\frac{N^{2}}{2}+\frac{\lambda^{4}}{8} N]$.  Thus, for an ohmic bath and finite operation time, correlations give a small correction to the usual distribution of probabilities.

\emph{Reducing the effects of long-range correlations---}It is well known that decoherence of a physical 
qubit can be systematically reduced by applying a series of NOT gates at a frequency \emph{higher} than the bosonic cut-off
\cite{VL98}. Although possible, the restriction $\omega \gg v_{b}\Lambda$ can be  experimentally very stringent.  The analysis here implies that essentially the same effect can be obtained for a logical qubit with $v_{b}\Lambda$ replaced by $\Delta^{-1}$.  For instance, a simple logical NOT, $\bar{X}=\sigma_{1}^{x}\sigma_{2}^{x}\sigma_{3}^{x}$, executed at half of each QEC cycle will change dramatically the effects of correlations in Steane's 3-qubit code.  Following the same steps as above, we now find that although the local probability of an error increases to $\approx3\epsilon/2$, the operator that captures the correlation between cycles is a \textit{second} derivative of the bosonic field.  Hence, Eq.~(\ref{OPE2}) would now read
\begin{equation}
\upsilon_{j=\left\{1,2,3\right\}}^{2}\sim3\epsilon/2+\left(\pi\lambda^{2}\Delta^{4}/32\right):\![\partial_{t}^{2}\theta]^{2}\label{RD}\!:,
\end{equation}
with $\upsilon_{0}^{2}=1-\sum_{j=\left\{1,2,3\right\}}\nu_{j}^{2}$.  The relevant two-point correlation function now decays faster, $\sim 1/t^8$, implying a much reduced effect of correlations. 

In the general case of an environment with spectral function $J(\omega)\sim \omega^{s}$, this simple change in the QEC code changes the decay of the relevant two-point correlation function from $\sim 1/t^{2(s+1)}$ to $\sim 1/t^{2(s+1+2n)}$, where $n$ is the number of logical NOTs in a cycle.  Such a modification may be overkill for the ohmic case. However, for sub-ohmic baths, $s<1$, this change in the QEC code tremendously improves its effectiveness against correlations.  Of particular importance is the fact that $1/f$ noise, so troublesome to qubits based on superconductors, can be thought of as the $s\to 0$ limit of the sub-ohmic case \cite{SMS02}.  Hence, we have shown that QEC codes can be improved so as to be effective in this case.

The effects of spatial correlations are also straightforward to discuss. The form of Eqs. (\ref{OPE1})-(\ref{RD}) remains the same; the order $\lambda^{2}$ terms are simply multiplied by an overall prefactor of order 1 which depends on the geometry. Thus, applying the logical NOT will reduce the effect of spatial correlations between logical qubits separated by a distance larger than $v_{\rm b}\Delta$. 

\newpage
\emph{Conclusions---}We have shown (1) that the dynamics of QEC introduces a natural scale into the problem of qubits coupled to a correlated environment, and (2) how to use this scale to decrease the computational error. For the spin-boson example, the result is a very useful combination of QEC with dynamical decoupling: QEC handles the short time decoherence while dynamical decoupling on the time scale established by QEC reduces the effect of long time correlations. 

We thank B. Terhal for helpful discussions.
This work was supported in part by (1) NSA and ARDA
under ARO contract DAAD19-02-1-0079 and (2) NSF Grant No. CCF-0523509.

\end{document}